\documentclass[a4paper,11pt]{article}
\usepackage{pos}

\usepackage{braket}
\usepackage{amsmath}
\usepackage{mathtools}
\usepackage{caption}
\usepackage{subcaption}
\usepackage{bm}
\usepackage{bbold}
\usepackage{xcolor}
\usepackage{tcolorbox}
\usepackage{amsfonts}
\usepackage{graphicx}
\usepackage{epstopdf}
\usepackage{multirow}
\usepackage[export]{adjustbox}
\usepackage[section]{placeins}

\numberwithin{figure}{section}

\title{\boldmath The lightest $D_0^\ast$ resonance from lattice QCD}

\ShortTitle{The lightest $D_0^\ast$ resonance from LQCD}

\author*{Nicolas Lang}

\affiliation{School of Mathematics and Hamilton Mathematics Institute, Trinity College, Dublin 2, Ireland}

\emailAdd{nicolas.lang@maths.tcd.ie}

\abstract{We recently presented elastic $I=1/2$ $D\pi$ scattering from lattice QCD at $m_{\pi} = 239$~MeV. The amplitude features a pole corresponding to a mass $m \approx 2200$~MeV and a width $\Gamma \approx 400$~MeV. The results were compared to an earlier study at a higher pion mass and to a similar study in the charm-strange sector. In this contribution to LATTICE2021 I summarize these results and compare them with experiment, based on the values reported by the particle data group. Our result lies significantly below the experimental $D_0^\ast$.  I also relate our findings to recent studies in chiral perturbation theory.\\
\vspace{0.5cm}\\
Based on work presented in \emph{JHEP2021(7),123} for the Hadron Spectrum Collaboration.}

\FullConference{%
 The 38th International Symposium on Lattice Field Theory, LATTICE2021
  26th-30th July, 2021
  Zoom/Gather@Massachusetts Institute of Technology
}

\newcommand{\nn}{\mathfrak{n}}

\definecolor{jlab_red}{RGB}{192,39,45}
\definecolor{jlab_orange}{RGB}{249,102,0}
\definecolor{jlab_blue}{RGB}{47,122,121}
\definecolor{jlab_green}{RGB}{65,125,10}

\definecolor{swave}{HTML}{B23B3B}
\definecolor{pwave}{HTML}{424874}
\definecolor{dstarpi}{HTML}{A15017} 

\begin{document}
\maketitle

\section{Introduction}
\begin{figure}[htp!]
	\centering
	\graphicspath{{plots/pdg_mass_plot/}}
	\resizebox{0.5\textwidth}{!}{\input{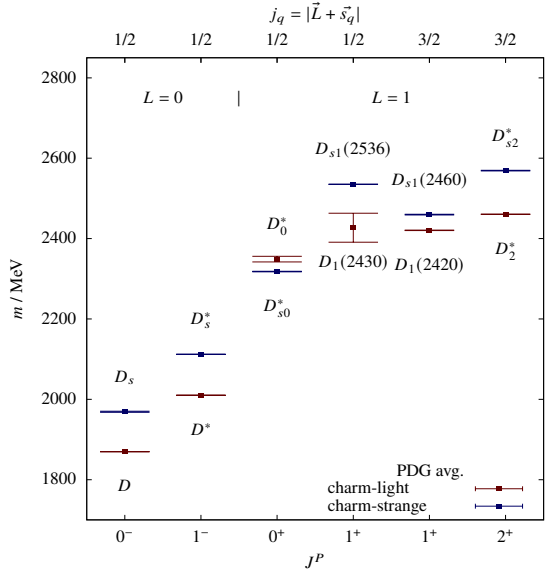}}
	\caption{Average mass values reported by the PDG~\cite{pdg:2018,pdg:2020} of the lowest charm-light and charm-strange states. The bottom horizontal axis shows the total angular momentum and parity $J^P$ whereas the top horizontal axis indicates the magnitude of the sum $j_q$ of the orbital angular momentum $\vec{L}$ and light-/strange-quark spin $\vec{s_q}$ of the corresponding quark-model state. The two quark spins are  conserved separately in the heavy-quark limit.}
	\label{fig:pdg}
\end{figure}

Open-charm systems are an interesting testing ground for our understanding of low-energy QCD.
The lightest scalar charm-light state  $D_0^\ast$ and its relation to the corresponding charm-strange state $D_{s0}^\ast$ has been a puzzle since the experimental discovery of the two mesons in 2003. From the perspective of the quark-model both states are represented by the scalar arising from a charm- and a light- or strange-quark interacting in a relative $P$-wave. The mass difference of the two states is therefore expected to be due, in large part, to the difference of the light and strange quark masses. However, the current particle data group (PDG) average~\cite{pdg:2018,pdg:2020} locates the $D_0^\ast$ at an energy compatible with that of the $D_{s0}^\ast$, as shown in fig. \ref{fig:pdg}. The large width of the $D_0^\ast$ that was found experimentally furthermore casts doubt on the validity of the quark model results for this system.

As a model independent approach, lattice QCD allows for the study of hadron resonances as they arise purely from QCD dynamics. While the current technical requirement to perform most calculations at larger-than-physical quark masses may be seen as a limitation, it can also be regarded as a tool to map out the quark mass dependence of the states between the flavour symmetric point and the physical mass.
In this contribution, I summarize the results from our recent study of $I=1/2$ $D\pi$ scattering~\cite{Lang:2021a}, which completes a quartet of studies of the scalar charm-light and charm-strange sector at two different mass points \cite{Moir:2016srx,Cheung:2020mql}. I also relate our results to recent studies in unitarised chiral perturbation theory and present an outlook for future investigations.

\section{Lattice Calculation}

The calculation is performed on an $(L/a_s)^3\times(T/a_t) = 32^3 \times 256$ anisotropic lattice with 2+1 dynamical quark flavours, $L$ and $T$ denoting the spatial and temporal lattice extents respectively, and $a_s$ ($a_t$) the spatial (temporal) lattice spacing. The anisotropy $\xi \equiv a_s/a_t \approx 3.5$ and the stable hadron masses are obtained from a fit of the relativistic dispersion relation
\begin{align}
	(a_t E)^2 = (a_t m)^2 + {\vec{d}}^2\left(\frac{2\pi}{\xi \;L/a_s}\right)^2
\end{align}
to the lattice energies of the respective hadron at different momenta. Here $\vec{d}$ is a vector of integers pointing in the momentum direction. We use the $\xi$ value from the pion fit in this calculation to convert energies to the rest-frame, but consider the D meson anisotropy in the assessment of systematic uncertainties. The scale setting is done through a comparison of the calculated $\Omega$ baryon mass on this ensemble with the physical one, such that $a_t^{-1} = m_\Omega^{\text{phys.}}/a_tm_\Omega^{\text{lat.}}$. The ensemble was generated using a tree-level Symanzik-improved anisotropic action in the gauge sector and a tree-level, tadpole-improved Sheikholeslami-Wohlert action for the fermions. The light quarks are heavier-than-physical while the strange quark is tuned to approximate the physical value. Table \ref{tab:ensem} summarises the most relevant properties of the ensemble.

\begin{table}[htb!]
	\centering
	\begin{tabular}{c|c|c|c|c|c}
		$a_s$ & $a_t^{-1}$ &  $(L/a_s)^3\times(T/a_t)$ & $m_{\pi}$ & 	$N_f $ & $N_{\text{cfg}} $ \\
		\hline
		 $0.11$ fm & 6.079 GeV & $32^3\times 256$ &$239$ MeV & $ 2 + 1$ & $ 484$ \\
		
	\end{tabular}
	\caption{Ensemble used in the calculation}
	\label{tab:ensem}
\end{table}

Correlators are computed using the distillation framework \cite{Peardon:2009gh}. All relevant graphs are evaluated, including disconnected pieces. The spectrum is obtained from a variational analysis of correlation functions obtained from a basis of interpolating operators, including $q\bar{q}$-like and meson-meson-like operators, solving the generalised eigenvalue equation
\begin{equation}
	C_{ij}(t) v_j^{(\nn)} = \lambda_\nn(t, t_0) C_{ij}(t_0) v_j^{(\nn)},   
\end{equation}
where $C_{ij}(t)$ is the matrix of correlators. Operators are projected to irreducible representations (\emph{irreps}) of the cubic group $O_h$ (at rest) and the little group $LG(\vec{p})$ at non-zero momentum to allow for the recovery of angular momentum information from the lattice spectra.
To obtain scattering amplitudes in the infinite volume we utilise the Lüscher quantisation condition~\cite{Luscher:1986pf,Luscher:1990ux,Luscher:1991cf} and its extensions~\cite{Rummukainen:1995vs,Kim:2005gf,Christ:2005gi,Fu:2011xz,Leskovec:2012gb,Hansen:2012tf,Briceno:2012yi,Guo:2012hv,Briceno:2014oea},
which allows a fit of a parametrised $\bm{t}$-matrix to the energy levels calculated in the finite volume. In our reference fit we parameterise the $\bm{t}$-matrix as
\begin{equation}
	\begin{aligned}
	(t^{(\ell)})^{-1}(s) &= \frac{1}{(2k)^\ell} K^{-1}(s) \frac{1}{(2k)^\ell} + I(s) \\
	K(s) &=  \frac{g^2} {m^2 - s} + \gamma \, .
	\end{aligned}
	\label{eq:kmatrix}
\end{equation}
$I(s)$ denotes the Chew-Mandelstam phase space subtracted at the energy of the pole parameter, $k(s)$ is the momenum in the center-of-momentum frame as a function of Mandelstam $s$ and $m$, $g$ and $\gamma$ are free parameters. To estimate the uncertainty arising from the choice of the specific functional form of the $t$ matrix, we fit a range of different parametrisations, among them different $K$-matrix forms, effective range and scattering length parametrisations as well as a Breit-Wigner. We also consider an amplitude based on unitarised chiral perturbation theory \cite{Guo:2018tjx}.

\section{Results}

\begin{figure}[htb!]
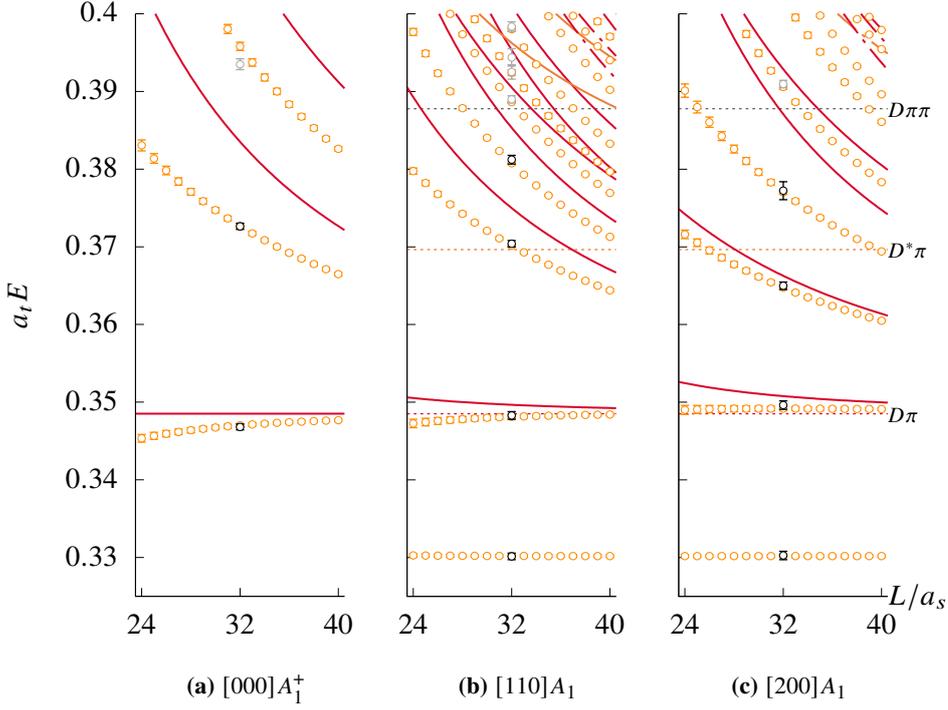

	\centering
	\begin{subfigure}[l]{0.23\textwidth}
		\hspace*{-1cm}
		\graphicspath{{plots/irreps/000_A1p/spectrum/}}
		\input{plots/irreps/000_A1p/spectrum/000_A1p_fvs_pub2.tex}
		\subcaption{$[000] A_1^+$}
	\end{subfigure}
	\begin{subfigure}[l]{0.23\textwidth}
		\hspace*{-1cm}
		\graphicspath{{plots/irreps/110_D2A1/spectrum/}}
		\input{plots/irreps/110_D2A1/spectrum/110_A1_fvs_pub2.tex}
		\subcaption{$[110] A_1$}
	\end{subfigure}
	\begin{subfigure}[l]{0.23\textwidth}
		\hspace*{-1cm}
		\graphicspath{{plots/irreps/200_D4A1/spectrum/}}
		\input{plots/irreps/200_D4A1/spectrum/200_A1_fvs_pub2.tex}
		\subcaption{$[200] A_1$}
	\end{subfigure}
	\caption{The finite volume spectrum in three of the irreps used to constrain the $D\pi$ S-wave amplitude. Black data points and error bars represent the energies obtained from the lattice. Dotted lines indicate thresholds. Solid lines correspond to non-interacting energies of operators included in the variational analysis. The solutions of the Lüscher determinant condition based on our reference amplitude parametrisation are superimposed in orange. (From fig. 5 and 6 in \cite{Lang:2021a})}
	\label{fig:fvs}
\end{figure}

We compute spectra for 10 irreps, both at rest and with one, two and three units of momentum. Figure \ref{fig:fvs} shows a subset of these.
We perform an $S$-wave fit based on 20 energy levels from 5 irreps where $l=0$ is the leading partial wave. There is a small but non-zero $P$-wave contribution in the moving-frame irreps, which we consider including a pole term in the $K$-matrix, accounting for an extra energy level far below $D\pi$ threshold appearing in these irreps. Using a single energy level from the $E^+$ irrep we show that the $D$-wave phase shift is consistent with zero and we therefore conclude that higher partial waves can be neglected. The energy levels used in the fit are all below the threshold where three-body $D\pi\pi$ scattering becomes kinematically possible and therefore, at this mass point, well below other thresholds of inelastic scattering. For the fit of our reference amplitude, using the $K$-matrix given by eq. \ref{eq:kmatrix} in the $S$ wave channel, we obtain $\chi^2/N_{\text{dof}} = \tfrac{13.49}{20 - 5} =  0.90$.

We analytically continue amplitudes to complex values of Mandelstam-$s$. Above $D \pi$ threshold there are two Riemann sheets due to the multi-particle branch cut, which are referred to as physical and unphysical sheet. Physical scattering occurs above the real axis on the physical sheet. Poles of the amplitude on the unphysical sheet may be interpreted as resonances and their residue gives the coupling to the decay channel. While the fits constrain the behaviour on the real axis, there may be large differences between parametrisations at complex energies. The amplitudes we consider are analytic, except for poles and the multi-particle branch cut along the real axis. Using the $K$-matrix formalism, they obey the constraint of unitarity by construction. We also aim to preserve causality by rejecting amplitudes that feature nearby poles above threshold on the physical sheet. The remaining amplitudes consistently feature an $S$ wave pole approximately $77 \pm 64$~MeV above threshold, with an imaginary part between $200$ and $600$~MeV, indicating a resonance as a universal feature.
For the pole and its residue we obtain the estimate
\begin{align*}
	\sqrt{s_0}\mathrm{/MeV} &= (2196 \pm 64) - \tfrac{i}{2} (425 \pm 224) \\
	c \mathrm{/MeV}        &= (1916 \pm 776)\exp i\pi( -0.59 \pm 0.41 )\;
\end{align*}
 from an envelope around the poles of all amplitude variations. We also take into account variations of the stable hadron masses and the anisotropy within their uncertainties, which are inputs to the Lüscher determinant condition. The amplitude (pole) resulting from our fits corresponds to the red curve (data point) in the left (right) panel of fig. \ref{fig:rhot_mass_dep}. The corresponding solutions of the Lüscher determinant condition yielding the finite volume energy levels are superimposed onto the spectrum in fig. \ref{fig:fvs}. 
\section{Interpretation}

Despite a heavier-than-physical pion mass, our result for the $D_0^\ast$ mass lies significantly below the experimental value reported by the PDG, which is approximately $2350$~MeV. It is interesting to note that the real part of the pole location is far from the point where the amplitude touches the unitarity bound (see fig. \ref{fig:rhot_mass_dep}). This difference is also reflected in our Breit-Wigner fit: whereas the pole of the Breit-Wigner amplitude is in agreement with our $K$ matrix result, the mass parameter of the amplitude takes on a value of $2380 \pm 36$~MeV, compatible with the PDG value.

\subsection*{Mass dependence}
\begin{figure}[htb!]
	\begin{center}
		\includegraphics[width=0.95\textwidth]{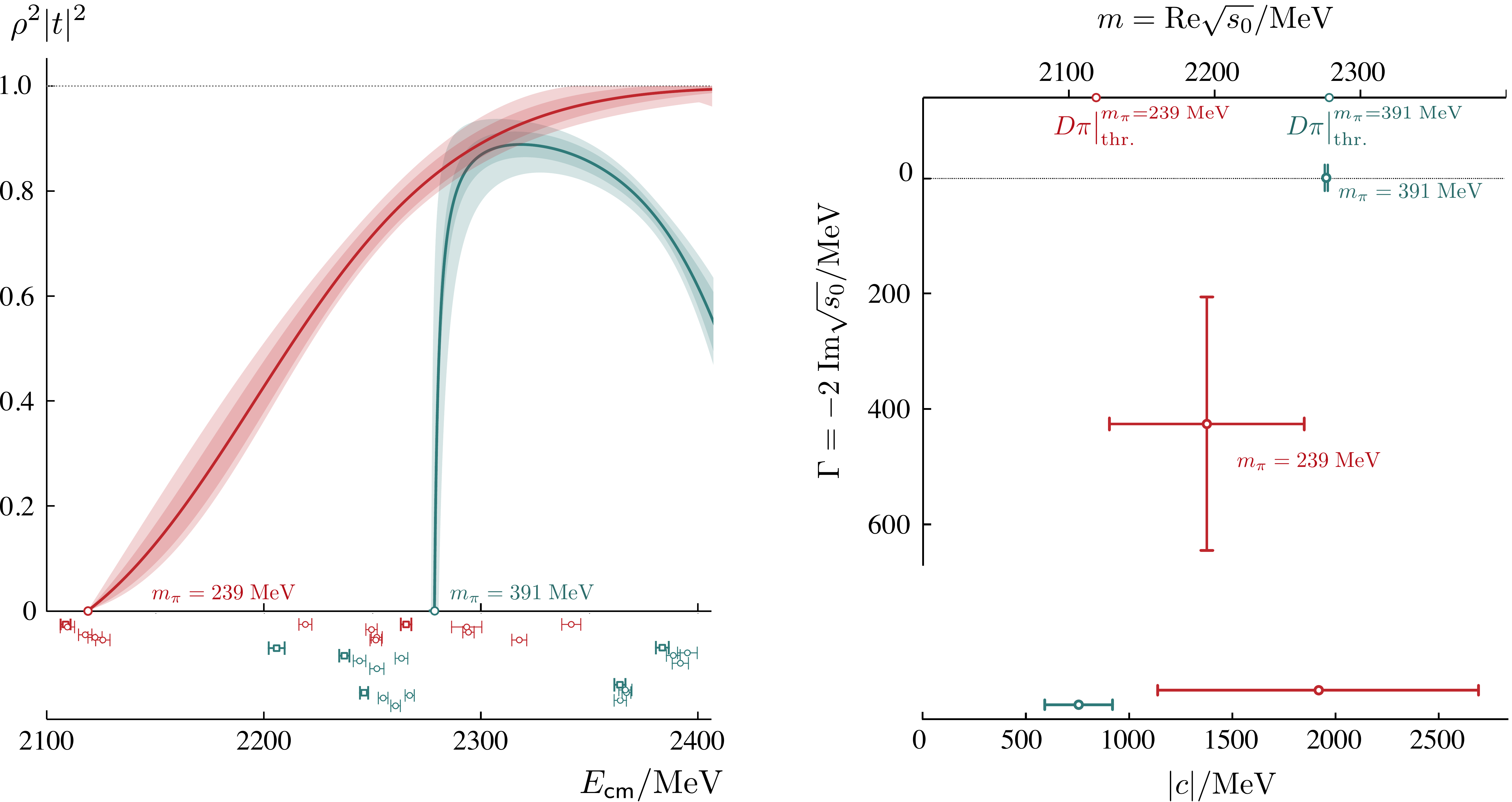}
		\caption{Comparison of the S-wave amplitudes (left panel), poles (right panel, top) and magnitude of couplings (right panel, bottom) at two different mass points. Energy levels constraining the amplitudes are indicated below the plot in the left panel. The inner bands around the amplitudes indicate the statistical uncertainty on the parameter values resulting from the fit. For $m_\pi=239$~MeV the outer band includes variations of the hadron masses, anisotropy and parametrisation, for $m_\pi=391$~MeV only variations of masses and anisotropy are included. (Fig. 10 in \cite{Lang:2021a})}
		\label{fig:rhot_mass_dep}
	\end{center}
\end{figure}

An earlier study of $D \pi$ scattering~\cite{Moir:2016srx} was conducted on an ensemble with a heavier pion mass $m_\pi=391$~MeV, whereas the ensemble used in our present study corresponds to $m_\pi=239$~MeV. This allows at least for the identification of a trend in the mass dependence of the amplitude. Fig. \ref{fig:rhot_mass_dep} shows the S-wave amplitudes and their poles obtained on both ensembles.
At the larger mass a shallow bound state was found just $2\pm1$~MeV below $D\pi$ threshold, which migrates into the complex plane at $m_\pi=239$~MeV. Both ampiltudes turn on rapidly above threshold indicating a large coupling to $D\pi$, which is reflected in the residue of the pole (see the bottom part of the right panel of fig. \ref{fig:rhot_mass_dep}).

\subsection*{Comparison to $DK$}
\begin{figure}[htb!]
	\centering
	\includegraphics[width=0.75\textwidth]{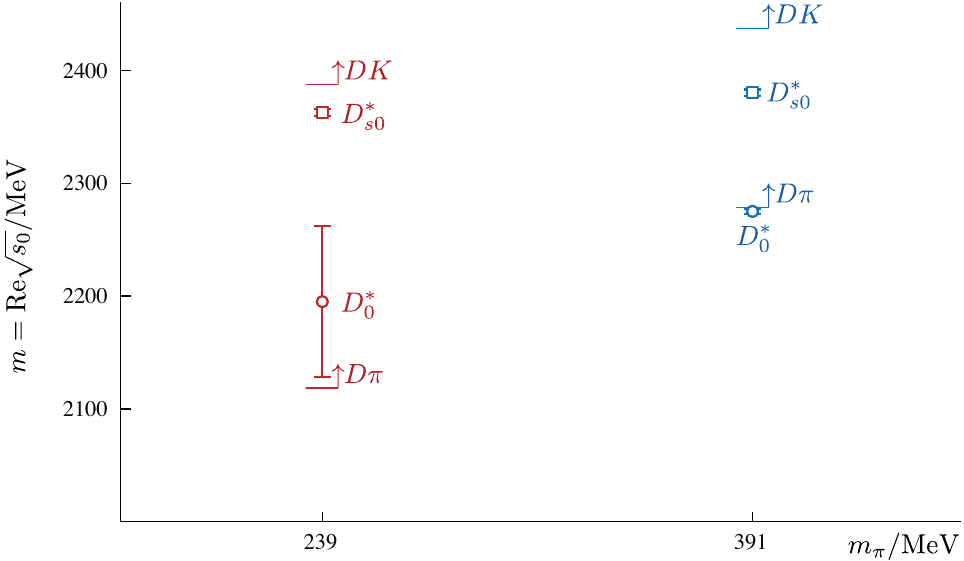}
	\caption{The real parts of the $S$-wave amplitude pole locations for $D\pi$ and $DK$ scattering at two different pion masses. The respective kinematic thresholds are indicated. (Fig. 13 in \cite{Lang:2021a})}

	\label{fig:pole_summary}
\end{figure}

As mentioned in the introduction, it is illuminating to compare the $D\pi$ system to the closely related $DK$ system. In \cite{Cheung:2020mql} $DK$ scattering amplitudes were calculated at both of the above-mentioned pion masses. 
We summarize the real parts of the poles extracted from the $S$-wave amplitudes at the two mass points for $I=1/2$ $D\pi$ and $I=0$ $DK$ in fig. \ref{fig:pole_summary}. The shallow $D\pi$ bound state at the higher pion mass evolves into a broad near-threshold resonance at the lower mass, while the $DK$ pole remains bound at both masses, showing only a very slight quark mass dependence. But clearly, the mass hiararchy of the two systems expected from the difference of the light and strange quark masses remains the same.
From the observed trend we would expect a resonant $D_0^{\ast}$ at the physical point whereas it is unclear whether the $D_{s0}^{\ast}$ would remain bound or evolve into a near-threshold resonance.

\subsection*{Comparison to $\chi_{\text{PT}}$ results}

In the limit where $m_u = m_d = m_s$ the $t$ matrix for open-charm scattering can be decomposed in terms of the irreps of $SU(3)_{\text{F}}$ as $ \bm{\overline{3}} \otimes \bm{8} = \bm{\overline{15}} \oplus \bm 6 \oplus \bm{\overline{3}}$. Using unitarised chiral perturbation theory, it was shown in \cite{Albaladejo:2017} that the anti-triplet pole splits into two when evolved in mass away from the $SU(3)$ limit towards the physical point. The mass evolution shown in Fig. 5 of the referenced work is roughly compatible with the poles obtained on our ensembles, when adjusting for the scale setting, suggesting that the $D_0^\ast$ and $D_{s0}^\ast$ correspond to the same $SU(3)_{\text{F}}$ multiplet. The mass evolution of the sextet state implies that the $D_0^\ast$ amplitude might feature an additional pole at a higher energy. The contribution by Guo et. al.~\cite{Guo:2021kdo} to LATTICE2021 makes a strong case for the existence of this pole. They perform a global fit including data obtained from the lattice for $m_{\pi} \approx 391$~MeV and predict the energy dependence of the phase shift for $m_{\pi} \approx 239$~MeV, which shows good agreement with our result close to $D\pi$ threshold.

\subsection*{Outlook}
Supporting the existence of the higher pole from the lattice side will require more data and the inclusion of the coupled channel in the scattering analysis. This could be an interesting future investigation. Additionally it would be worthwhile to perform this study at further pion masses, providing constraints on the mass evolution predicted by chiral perturbation theory. Lastly, a similar set of open-charm axial-vector states is predicted and a corresponding analysis performed in the $D^\ast \pi$ and $D^\ast K$ channels would allow for a comparison.

\section*{Acknowledgements}
I would like to thank David Wilson and Christopher Thomas for helpful comments.

\FloatBarrier

\bibliography{biblio}{}
\bibliographystyle{unsrt}

\end{document}